\documentclass[11pt,twoside]{article}

\usepackage{asp2014}

\aspSuppressVolSlug
\resetcounters

\begin{document}

\title{Interoperable geographically distributed astronomical infrastructures: technical solutions.}

\author{S.~Bertocco,$^1$ B.~Major,$^2$ P.Dowler,$^2$ S.Gaudet,$^2$ G.Taffoni,$^1$ and F.Pasian$^1$}
\affil{$^1$, INAF-OATs, Istituto Nazionale di Astrofisica - Osservatorio Astronomico di Trieste, Trieste, Italy}
\affil{$^2$CADC, National Research Council Canada, Victoria, BC, Canada}

\paperauthor{Sara~Bertocco}{bertocco@oats.inaf.it}{orcid.org/0000-0003-2386-623X}{Istituto Nazionale di Astrofisica}{Osservatorio Astronomico di Trieste}{Trieste}{}{34143}{Italy}
\paperauthor{Brian~Major}{majorb@nrc-cnrc.gc.ca}{0000-0001-9263-3048}{National Research Council Canada}{CADC}{Victoria}{British Columbia}{}{Canada}
\paperauthor{Patrick~Dowler}{patrick.dowler@nrc-cnrc.gc.ca}{orcid.org/0000-0001-7011-4589}{National Research Council Canada}{CADC}{Victoria}{British Columbia}{}{Canada}
\paperauthor{Séverin~Gaudet}{Severin.Gaudet@nrc-cnrc.gc.ca}{orcid.org/0000-0001-9281-2361}{National Research Council Canada}{CADC}{Victoria}{British Columbia}{}{Canada}
\paperauthor{Fabio~Pasian}{pasian@oats.inaf.it}{orcid.org/0000-0002-4869-3227}{Istituto Nazionale di Astrofisica}{Osservatorio Astronomico di Trieste}{Trieste}{}{34143}{Italy}
\paperauthor{Giuliano~Taffoni}{taffoni@oats.inaf.it}{orcid.org/0000-0002-4211-6816}{Istituto Nazionale di Astrofisica}{Osservatorio Astronomico di Trieste}{Trieste}{}{34143}{Italy}

\begin{abstract}
The increase of astronomical data produced by a new generation of observational tools poses the need to distribute data and to bring computation close to the data. Trying to answer this need, we set up a  federated data and computing infrastructure involving an international cloud facility, EGI federated, and a set of services implementing  IVOA standards and recommendations for authentication, data sharing and resource access.
In this paper we describe technical problems faced, specifically we show the designing, technological and architectural solutions adopted. We depict our technological overall solution to bring data close to computation resources. Besides the adopted solutions, we propose some points for an open discussion on authentication and authorization mechanisms.

%
%

\end{abstract}

\section{Introduction}
In recent years, the growing amount of data coming from the new generation instruments boosted in the Astronomy and Astrophysics community the need to distribute data and to build distributed infrastructures to  share data and to bring computation  close to the data. In a joint project between the Canadian Advanced Network for Astronomical Research (CANFAR) and the INAF-Osservatorio Astronomico di Trieste (OATs), partially funded by the EGI-Engage H2020 European Project, we built a federated international cloud facility  making available storage and computation resources powered by astronomical specific tools for data sharing and computation and  transparently interoperable with other analogous software infrastructures geographically distributed. For this purpose, at OATs-INAF have been deployed a set of astronomical services, based on an IVOA standards and recommendations and open-source software released by CADC (Canadian Astronomical Data Center)\citet{opencadcrepo}. These services are provided as cloud resources through an OpenStack-based cloud site, deployed at  OATs-INAF, and interoperable with IVOA-based services hosted by CANFAR (Canadian Advanced Network for Astronomy Research). This cloud site is furthermore federated in the EGI FedCloud, allowing to the EGI users to access the  Astronomical and Astrophysical specific resources hosted both at OATs-INAF and CANFAR.

%
%

\section{The overall architecture}
\begin{figure}[ht]
\centering\includegraphics[width=1.1\linewidth]{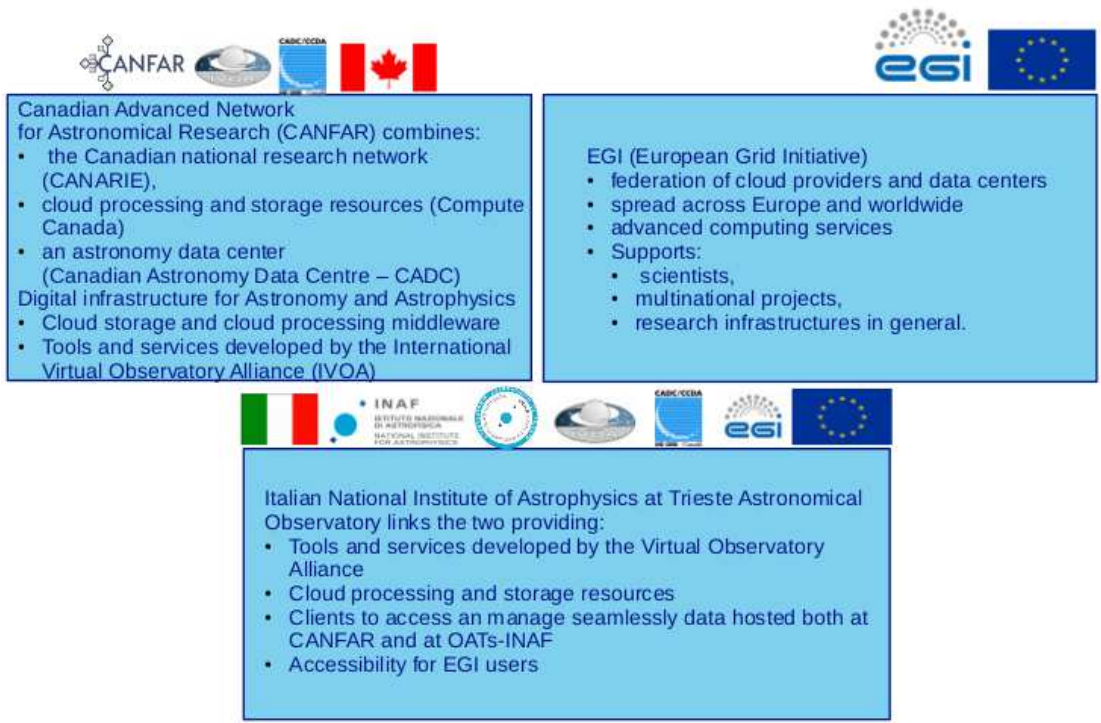}
\caption{Main architectural components}\label{fig:architecture}
\end{figure}

The overall architecture is composed by three main building blocks: (i) the CANFAR infrastructure hosting cloud processing and storage resources, an astronomy data center and tools and services using International Virtual Observatory Alliance (IVOA) standards; 
(ii) the OATs-INAF infrastructure hosting cloud processing and storage resources, control access and data management tools and services based on IVOA standards and recommendations, clients to seamlessly access and manage data hosted both at CANFAR and at OATs-INAF; (iii) the EGI FederatedCloud, which is a federation of cloud providers and data centers spread across Europe and worldwide providing advanced computing services to support scientists, multinational projects and research infrastructures in general (see figure~\ref{fig:architecture}).
This paper is focused on the work done on the OATs-INAF infrastructure which establishes a link between the CANFAR infrastructure and the EGI FedCloud. It is federated in EGI with its OpenStack-based cloud site, providing EGI users with access to his cloud resources, and it offers client tools to access and manage data hosted both at OATs-INAF and at CANFAR thanks to the IVOA standard and recommendation implementation.

\section{Access Control and Storage services IVOA standards and recommendations based}

The Canadian Advanced Network for Astronomical Research (CANFAR) provides  its users with access to  large resources for both storage and processing, using a cloud based framework. The available services are based on open-source software libraries, released by CADC, implementing IVOA standards and recommendations. At OATs-INAF a twin framework has been deployed, with the same technical requirements and use cases, based on the same software libraries, covering user access and data management aspects~\citep{P6-078_adassxxvii}.

\textbf{Users Access Control:} The user access is managed by an Access Control Service. It is a RESTful service, Java written, exploiting the Restlet API and the Oracle JAAS (Java Authentication and Authorization Service). This service allows users registration and authentication and user groups management. For each user multiple identities are supported (at present X.509 certificate, user/password and cookies are implemented, but they can be easily extended). The information about users, groups and group memberships is stored in an LDAP server. Source code is available at \citet{opencadcrepo}
 
\textbf{Authorization:} Users are authorized to a resource (e.g. a service or proprietary data) if they are a member of the group(s) protecting that resource. Access rights to a specific resource are granted by the resource owner that is also in charge to assign  group membership  to other users.  For example, if user A, owner of the resource X, wants grant  to user B access to this resource, he must create group G, add user B to group G and grant access to group G to the resource X.
The information about access rights is stored as meta-data of the resource in a relational database.

\textbf{Storage service}
The storage service implements the VOSpace IVOA recommendation \citet{ivoaVOSpace}: "A VOSpace web service is an access point for a distributed storage network. Through this access point, a client can: add or delete data objects, manipulate metadata for the data object, obtain URIs through which the content of the data objects can be accessed. VOSpace does not define how the data is stored or transferred, but only the control messages to gain access. Thus, the VOSpace interface can readily be added to an existing storage system."
Given this definition, an IVOA-based storage service needs three components: (i) a VOSpace interface, responsible for the meta-data management; (ii) a data transfer service, responsible for handling the information on how to upload and download data; (iii) a a storage service management to connect a VOSpace interface implementation with a specific storage solution, in charge to manage the physical storage layer and to point to the data locations.
The OATs-INAF Storage services is implemented using the CADC open-source software libraries\citet{opencadcrepo} for the VOSpace interface implementation but OATs-INAF designed and developed the component for the transfer service and the storage management\citet{oatscadcrepo}. It is a Java RESTful web service, realized exploiting the Restlet APIs, persisting meta-data in a relational database. Mysql, MariaDB, Sybase and Postgres have been tested and used  both in testing and production environment. The data access is managed reading f the information about the groups having access from the database. The group name is an identifier containing the identifier of the authority hosting the group. The querying user's identity is verified through the access control service registered by the authority hosting the group and, when the user is recognized, his credentials are taken by the local credential delegation service and used to access the data. 

\section{OATs-INAF cloud site}
An OpenStack-based cloud site has been deployed at OATs-INAF with three main purposes: to provide a computational and storage resources infrastructure to OATs-INAF users; to use this infrastructure to access the IVOA-based geographically distributed storage services, at present both at OATs-INAF and CANFAR; to register this site in the EGI FedCloud and to use it as cross link between EGI and the CANFAR infrastructure, giving to European users transparent access to CANFAR hosted data.

\section{Interoperability}
The interoperability of services and infrastructures is based on the use of X.509 personal certificates.  
CANFAR and OATs-INAF Access Control services maintain almost three identities of a registered user: a numeric identity (for internal purpose), a username/password identity and a X.509 certificate identity. 
EGI supports X.509 authentication and authorization through VOMS and Keystone-VOMS module in the OpenStack-based cloud. This way, each user having a X.509 personal certificate and member of a Virtual Organization can create a VOMS proxy certificate allowing him to access the EGI FedCloud site at  OATs-INAF and the IVOA based services at OATs-INAF and CANFAR. The OATs-INAF cloud site, federated in the EGI FedCloud, provides users with a virtual machine image conveniently configured to allow them to log-in with the same credentials used to access the IVOA-based services at OATs-INAF. This machine contains clients to access both OATs-INAF and CANFAR services like VOSpace and a set of tools specific for data reduction based on ESO Scisoft.

\section{Future plans}
The work done  is the base for future developments in different areas. The IVOA based services deployability can be made easier providing containers with pre-installed services. The user authentication can be integrated adding  identity federation services (e.g. EduGAIN) and making the user's credential delegation token based. At present the group membership verification affecting the authorization  is strongly linked on the hypothesis that the group name is an identifier containing the authority hosting the group. Some work has to be done in the IVOA community to find consensus on this solution or to study a different one and possibly to release a standard on authorization solutions.

\bibliography{P6-078}  

\end{document}